\documentstyle[aps,prl,twocolumn,floats,epsf]{revtex}
\newcommand{\ph}{{\phantom{\dagger}}}
\begin{document}
\draft
\title{Density of states near the Mott--Hubbard transition 
in the limit of large dimensions}
\author{Stefan Kehrein}
\address{Theoretische Physik III, Elektronische Korrelationen und
Magnetismus,\\
Institut f\"ur Physik, Universit\"at~Augsburg, 86135~Augsburg, Germany \\
{\rm(August 26, 1998)}
}

\address{~
\parbox{15cm}{\rm
\medskip
The zero temperature Mott--Hubbard transition as a function of
the Coulomb repulsion $U$ is investigated
in the limit of large dimensions. The behavior of the density of 
states near the transition at $U=U_c$
is analyzed in all orders of the skeleton expansion.
It is shown that only two transition scenarios are consistent with the
skeleton expansion for~$U<U_c$: (i)~The Mott--Hubbard transition is 
``discontinuous'' in the sense that in the density of states 
finite spectral weight is redistributed at~$U_c$. (ii)~The transition
occurs via a point at $U=U_c$ where the system 
is neither a Fermi liquid nor an insulator. 
\\~\\
PACS numbers: 71.30.+h, 71.27.+a, 71.28.+d
}}

\maketitle

\narrowtext
The correlation--induced metal--insulator transition \cite{Mott61}
is a fundamental and challenging problem in condensed matter physics.
Theoretical work has centered around the Hubbard model, where a
metal--insulator transition is expected at half--filling for some 
critical on--site Coulomb repulsion~$U_c$. A new line of
approach for studying the metal--insulator transition has recently 
been opened by the limit of high dimensions 
\cite{Metzner89}. At present this
{\em dynamical mean field theory} seems the only tractable method 
to obtain exact statements concerning the Mott--Hubbard
transition. Such results should be of considerable interest
also for the physically relevant threedimensional case.

In the dynamical mean field approach the behavior of the density of states 
$\rho(\epsilon)$ near the Mott--Hubbard transition at half--filling and 
zero temperature 
is of particular interest. Georges et al. \cite{RMP96} have described
in detail a transition scenario where the spectral
weight in the vicinity of the Fermi surface 
vanishes continuously as~$U\uparrow U_c$ (compare Fig.~1):
In some finite interval 
$[\epsilon_F-\Delta,\epsilon_F+\Delta]$ around the Fermi surface one finds
\begin{equation}
\int_{\epsilon_F-\Delta}^{\epsilon_F+\Delta} \rho(\epsilon)
\longrightarrow 0 \qquad \mbox{as~} U\uparrow U_c \ .
\label{scenario1}
\end{equation}
This implies that at~$U_c$ a finite excitation gap of size~$2\Delta$ opens
in the density of states turning the Fermi liquid at once into an insulator.
However, the analytical solution of the $d\rightarrow\infty$
Hubbard model is unfortunately still out of reach and the above transition
scenario has been discussed controversially in the literature 
(see e.g. Refs.~\cite{Gebhard,Logan98}). 

The purpose of this Letter is to establish constraints regarding possible
transition scenarios, based on an analysis of the dynamical mean field
equations in all orders of the skeleton  expansion.
Conceptually the arguments are similar to the 
reasoning used by Luttinger \cite{Luttinger61} and Langer \cite{Langer61}
in order to derive the $(\epsilon-\epsilon_F)^2$--behavior of
the imaginary part of the self--energy in the vicinity of the 
Fermi surface. An asymptotic
sum rule for the imaginary part of the self--energy can be established
that implies the following constraint for the transition scenario:
For all intervals with $\Delta>0$ around the Fermi surface
the limit in Eq.~(\ref{scenario1}) has to be nonzero as $U\uparrow U_c$.
Therefore the Mott--Hubbard transition investigated here is
either (i) discontinuous in the sense that finite spectral weight is
redistributed at~$U_c$, or (ii) there is nonzero spectral weight in any
neighborhood of the Fermi surface for~$U=U_c$. 

Possibility (ii) 
implies that the density of states is still gapless at the transition,
i.e.\ pseudogap--like with $\rho(\epsilon_F)=0$ but 
$\rho(\epsilon)\propto |\epsilon-\epsilon_F|^\alpha$, $\alpha>0$ in the
vicinity of~$\epsilon_F$. This
would signal the existence of a non--Fermi liquid point
separating metallic and insulating regimes.
The analysis in this Letter will also provide insights
in various analytical approximation schemes devised
to solve the dynamical mean field equations.

The Hamiltonian of the Hubbard model is
\[
H=-\frac{t}{\sqrt{d}}\sum_{(ij),\alpha} c^\dag_{i\alpha} c^\ph_{j\alpha}
+U\sum_i \Big(c^\dag_{i\uparrow} c^\ph_{i\uparrow}-\frac{1}{2}\Big)
\Big(c^\dag_{i\downarrow} c^\ph_{i\downarrow}-\frac{1}{2}\Big) 
\]
where the hopping matrix elements are scaled as 
$t_{ij}\rightarrow t/\sqrt{d}$ to obtain a physically meaningful
limit of large dimensions \cite{Metzner89}. In the sequel $t$ sets 
the energy scale and half--filling corresponds to $\epsilon_F=0$.
For $d\rightarrow\infty$ an effective action describing
the single--site dynamics of one fermionic degree of freedom 
$(c^\ph_{0\alpha},c^\dag_{0\alpha})$ can be written as
(for details see Ref.~\cite{RMP96})
\begin{eqnarray}
S_{\rm eff}&=&-\int_0^\beta d\tau\,d\tau' \sum_\alpha
c^\dag_{0\alpha}(\tau) {\cal G}^{-1}(\tau-\tau') c^\ph_{0\alpha}(\tau') 
\label{eff_action} \\
&+&\! U\int_0^\beta d\tau 
\left(c^\dag_{0\uparrow}(\tau) c^\ph_{0\uparrow}(\tau)-\frac{1}{2}\right)
\left(c^\dag_{0\downarrow}(\tau) c^\ph_{0\downarrow}(\tau)-\frac{1}{2}
\right) .
\nonumber
\end{eqnarray} 
All local correlation functions of the original Hubbard model can be
derived from (\ref{eff_action}). The effective action is supplemented 
by a self--consistency
condition relating the Weiss local field ${\cal G}(\tau-\tau')$ and
the local propagator \linebreak
$G(\tau-\tau')=-\langle Tc_0(\tau)c_0^\dag(\tau')\rangle$.

We will investigate the self--consistency problem
by using a Bethe lattice with large coordination number. 
On a Bethe lattice the self--consistency condition takes
a particularly simple form~\cite{RMP96}
\begin{equation}
{\cal G}^{-1}(i\omega_n)=i\omega_n-t^2\,G(i\omega_n) \ .
\label{Bethe_sc}
\end{equation}
The discussion will be restricted to zero temperature, half--filling 
and the {\em paramagnetic phase}. Solutions with magnetic ordering
are therefore excluded \cite{lro}.

The $k$--independent self--energy of the Hubbard model 
is given by $\Sigma(\epsilon^+)={\cal G}^{-1}(\epsilon^+)-G^{-1}(\epsilon^+)$
or
\[
\Sigma(\epsilon^+)=\epsilon^+-t^2\int_{-\infty}^\infty d\omega
\frac{\rho(\omega)}{\epsilon^+-\omega} 
-\left(\int_{-\infty}^\infty d\omega
\frac{\rho(\omega)}{\epsilon^+-\omega}\right)^{-1} 
\]
where $\rho(\epsilon)=-\frac{1}{\pi}\mbox{Im}\, G(\epsilon^+)$ is the local
density of states and $\epsilon^+=\epsilon+i0^+$. 
By introducing the principal value integral
$\Lambda(\epsilon)=\mbox{P}\int_{-\infty}^\infty d\omega\,
\frac{\rho(\omega)}{\epsilon-\omega}$ one can split the self--energy
into its real and imaginary parts 
$\Sigma(\epsilon^+)=K(\epsilon)-iJ(\epsilon)$ 
leading to
\begin{eqnarray}
K(\epsilon)&=&\epsilon-t^2\Lambda(\epsilon)
-\frac{\Lambda(\epsilon)}{\Lambda(\epsilon)^2+\pi^2\rho(\epsilon)^2}
\label{defK} \\
J(\epsilon)&=&-\pi t^2\rho(\epsilon)
+\frac{\pi\rho(\epsilon)}{\Lambda(\epsilon)^2+\pi^2\rho(\epsilon)^2} \ \ .
\label{defJ}
\end{eqnarray}
Notice
that $\rho(\epsilon)$ is symmetric and therefore $\Lambda(\epsilon)$
antisymmetric at half--filling. 

We investigate solutions near the metal--insulator
transition coming from the metallic side $U\uparrow U_c$ \cite{Uc1Uc2}.
A key role in 
this transition in the limit of large dimensions is played by the 
fixed value of the density of states at the Fermi surface
$\rho(\epsilon_F)=1/\pi t$ for the half--filled case  
for $U<U_c$ \cite{MuellerHartmann89}. This is due to 
the Fermi liquid property $J(\epsilon_F)=0$. In the zero temperature 
metal--insulator transition scenario described by Georges et al. 
\cite{RMP96} the spectral weight~$w$ 
in the vicinity of the Fermi surface vanishes continuously
while still $\rho(\epsilon_F)=1/\pi t$ for~$U<U_c$.
Rephrased mathematically,   
the local density of states separates in a low-- and a
high--energy part for $U\uparrow U_c$
\begin{equation}
\rho(\epsilon)=\rho_l(\epsilon)+\rho_h(\epsilon) \ ,
\label{energysep}
\end{equation}
where $\rho_h(\epsilon)$ vanishes in some finite interval
$[-\Delta,\Delta]$ around the Fermi surface. The low--energy part
$\rho_l(\epsilon)$ describes a quasiparticle resonance at the
Fermi surface with a continuously vanishing spectral weight~$w$
as \linebreak $U\uparrow U_c$. Eq.~(\ref{energysep})
is identical to the ansatz of Moeller et al. \cite{Moeller95},
see also Fig.~1.

Due to the pinning $\rho_l(\epsilon_F)=1/\pi t$ the quasiparticle
resonance at the Fermi surface has a width of order~$wt$. 
The existence of one single low--energy scale~$wt$ in this transition
scenario implies that the following scaling ansatz becomes possible in the 
limit $w\rightarrow 0$ \cite{Moeller95}
\begin{equation}
\rho_l(\epsilon)=\frac{1}{t}\:f\left(\frac{\epsilon}{wt}\right) \ . 
\label{scaling}
\end{equation}
The dimensionless function~$f(x)$ is normalized, \linebreak
$\int_{-\infty}^\infty dx\,f(x)=1$, and fulfills
$f(0)=1/\pi$. This implies
\begin{equation}
\int_{-\Delta}^{\Delta} d\epsilon\,\rho(\epsilon)=w\rightarrow 0 
\quad\mbox{for~} U\uparrow U_c \ . 
\label{vanishing_w}
\end{equation}
$\rho_h(\epsilon)$ also
has some $w$--dependence since it contains the remaining spectral
weight~$1-w$. This weak dependence will be 
of no importance in the following discussion.

\begin{figure}[t]
\begin{center}
\leavevmode
\epsfxsize=8.5cm
\epsfysize=6cm
\epsfbox{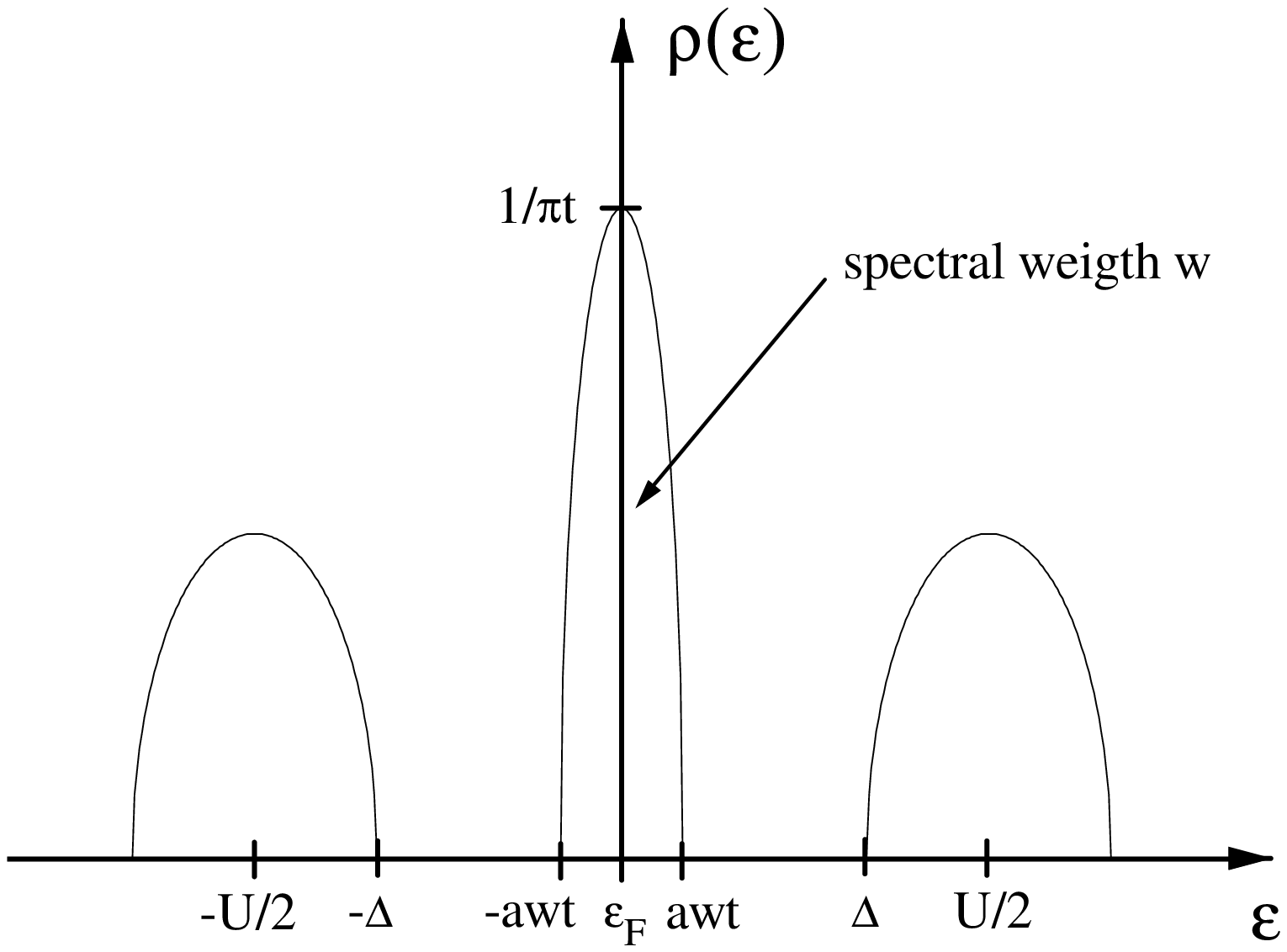}
\vspace*{-0.5cm}
\caption{Hypothetical density of states near a metal--insulator 
transition with vanishing spectral weight in the vicinity of the
Fermi surface.}
\end{center}
\end{figure}

We now show that the above transition 
scenario cannot be realized within the skeleton expansion. 
It is taken for granted here that the transition occurs at 
a finite critical~$U_c<\infty$ as supported by numerical
calculations for nonzero temperature (see Ref.~\cite{RMP96}). 

Before proceeding with actual calculations we present an
intuitive argument 
why the scenario~(\ref{energysep}) is not consistent with
the skeleton expansion.
Consider a (hypothetical) density of
states near a metal--insulator transition as
sketched in Fig.~1. For simplicity we assume that $f(x)$ has
compact support in $[-a,a]$. Then 
for sufficiently small~$w$ the principal value integral 
$\Lambda(\epsilon)$ develops zeroes for $\epsilon=\pm\epsilon_0$ 
with $\epsilon_0$ of order~$\sqrt{w}t$. Obviously 
$awt\ll\epsilon_0\ll \Delta$ and therefore $\epsilon_0$
lies in the gap $\rho(\pm\epsilon_0)=0$ in the limit~$w\rightarrow 0$.
In this case  
$\Lambda(\epsilon_0)=0$ leads to a $\delta$--function--like
contribution in $J(\epsilon)$ according
to Eqs.~(\ref{defK}) and (\ref{defJ}). 
Hence a density of states as depicted in Fig.~1
enforces an imaginary part of the self--energy that vanishes
everywhere in the gap {\em except} for $\delta$--functions 
at~$\pm\epsilon_0$ \cite{ZRK93} (notice that according to Eq.~(\ref{defJ}) 
$J(\epsilon)$ is well--behaved for all other energies 
including the band edges). In the 
skeleton expansion the imaginary part of the self--energy is related 
to the available phase space for scattering processes. 
With the above ansatz for $\rho(\epsilon)$, however, 
this phase space does not increase when passing from 
the small energy interval $[-awt,awt]$ to
$[-\Delta,\Delta]$. Therefore the resonance of $J(\epsilon)$ in the gap
cannot be explained in any order of the skeleton expansion.
This argument can be generalized also to situations without
a true gap.

Let us formalize this reasoning. 
The method used is a demonstration by contradiction, i.e.\ we first
assume that solutions of the $d\rightarrow\infty$ self--consistency
conditions with the property~(\ref{vanishing_w}) exist. In the sequel 
$f(x)$ need not have compact support. Consider
the value of $\Lambda(\Delta)$ in the limit~$w\rightarrow 0$. The
contribution from the states $|\epsilon|<\Delta$ in the principal value
integral for $\Lambda(\Delta)$ behaves as 
\[
\int_{-\Delta}^{\Delta}d\epsilon\,\frac{\rho(\epsilon)}{\Delta-\epsilon}
\approx\frac{1}{\Delta}\int_{-\Delta}^{\Delta}d\epsilon\,\rho(\epsilon)
=w/\Delta \ , 
\]
whereas
the remaining spectral weight of order~$w^0$ for $|\epsilon|>\Delta$
contributes an essentially $w$--independent negative term. 
Therefore eventually $\Lambda(\Delta)<0$, and thereby, according to
(\ref{defK}) 
\begin{equation}
K(\Delta)>\Delta \quad\mbox{for~} w\rightarrow 0 \ .
\label{K>}
\end{equation}
It will be demonstrated below that the skeleton expansion
implies $K(\Delta)\leq 0$, yielding a contradiction. In order
to show this we derive a sum rule for the
imaginary part of the self--energy. 
The self--energy can be expressed in the 
following manner in a skeleton expansion~\cite{LuttingerWard60} 
\begin{eqnarray}
\Sigma(z)&=&\big[ \mbox{all possible skeleton diagrams with the} 
\nonumber \\
&&\mbox{unperturbed propagator~}{\cal G}(z)\mbox{~replaced}
\nonumber \\
&&\mbox{by the true propagator~}G(z) \big] \ .
\label{skeleton}
\end{eqnarray}
The diagrams in this series can be
analyzed in an elegant way introduced by Langer \cite{Langer61}:
The imaginary part of a skeleton diagram is given
by all possible Cutkosky--cuts across the internal lines. 
Diagrams with cuts across $n$~internal hole and 
$n+1$ internal particle lines contribute
\begin{eqnarray}
&&\int_0^\infty d\omega_1\ldots d\omega_{n+1} \:
\delta\left(\sum_{i=1}^{n+1}\omega_i-\epsilon\right)
\label{skel_n} \\
&\times&\prod_{i=1}^n \left[ \int_0^{\omega_i} d\xi_i
\,\rho(\xi_i)\rho(\xi_i-\omega_i) \right] 
\:\times\rho(\omega_{n+1}) \nonumber\\
&\times&\Gamma^{(n)}(\xi_1,\xi_1-\omega_1,\xi_2,\xi_2-\omega_2,\ldots,
\omega_{n+1}) \nonumber 
\end{eqnarray} 
to $J(\epsilon)$,
where $\Gamma^{(n)}$ describes generalized vertex functions \cite{Langer61}.
In the vicinity of the Fermi surface this expansion 
gives the well--known behavior of the imaginary part of the
self--energy for $\epsilon\rightarrow\epsilon_F$ \cite{Luttinger61}
\[
J(\epsilon)=\sum_{n=1}^\infty \frac{\Gamma_n \, \rho(\epsilon_F)^{2n+1}}{
(2n)!} \: (\epsilon-\epsilon_F)^{2n} \ ,
\]
where the coefficients $\Gamma_n$ are given by the above 
vertex functions and their
derivatives in the infrared limit.
The reason behind the increasing powers of~$\epsilon$
is the restriction of the available phase space for scattering
processes in (\ref{skel_n}) in the limit~$\epsilon\rightarrow 0$. 

Next we investigate the behavior of 
$\int_{-\Delta}^{\Delta} d\epsilon\,J(\epsilon)$ in the limit $w\rightarrow 0$.
The key observation is that the ansatz~(\ref{vanishing_w})
{\em automatically} leads to a restriction of the available phase
space for scattering processes on an energy scale smaller than~$\Delta$:
We can consider the limit $w\rightarrow 0$ in (\ref{skel_n}) and 
this leads to an expression for the integrated~$J(\epsilon)$
\begin{equation}
\int_{-\Delta}^{\Delta} d\epsilon\,J(\epsilon) = \sum_{n=1}^\infty
\tilde\Gamma_n \, \left(\frac{w}{2}\right)^{2n+1} \ .
\label{expansion_intJ}
\end{equation}
The coefficients $\tilde\Gamma_n$ are averages of the
functions $\Gamma^{(n)}(\xi_1,\xi_1-\omega_1,\xi_2,\xi_2-\omega_2,\ldots,
\omega_{n+1})$ over the Fermi liquid region: In (\ref{skel_n}) the vertex 
functions are only probed on the Fermi liquid energy scale of order~$wt$ in
the integral $\int_{-\Delta}^{\Delta} d\epsilon\,J(\epsilon)$. Therefore
one can infer the scaling behavior of $\tilde\Gamma_n$ with~$w$
from the behavior in the infrared limit: $\tilde\Gamma_n\propto\Gamma_n$
for $w\rightarrow 0$ \cite{uniformity}. On the other hand
from the scaling ansatz~(\ref{scaling}) one
deduces $\Gamma_n\propto w^{-2n}$ with an $w$--independent
proportionality constant depending on the function~$f(x)$. 
Therefore the terms in the series~(\ref{expansion_intJ})
are of order~$w$ leading to
\begin{equation}
\int_{-\Delta}^{\Delta} d\epsilon\,J(\epsilon)= \alpha\,w\,t^2+O(w^2)
\label{sumrule}
\end{equation}
with some dimensionless constant~$\alpha$ that is finite for 
an integrable function $f(x)$.
Since $K(\epsilon)$ and $J(\epsilon)$ are connected by a 
Kramers--Kronig relation \cite{Luttinger61}
\begin{equation}
K(\epsilon)=\frac{1}{\pi} \:\mbox{P}\int_{-\infty}^\infty
d\omega\,\frac{J(\omega)}{\epsilon-\omega} \ ,
\label{kramerskronig}
\end{equation}
Eq.~(\ref{sumrule}) implies that the positive contributions to $K(\Delta)$
in~(\ref{kramerskronig}) vanish for $w\rightarrow 0$,
leading to
\begin{equation}
K(\Delta)\leq 0 \quad\mbox{for~} w\rightarrow 0 \ .
\label{Kleq}
\end{equation}
Eqs.~(\ref{K>}) and (\ref{Kleq}) are in obvious contradiction in
the limit~$w\rightarrow 0$: For sufficiently small but 
{\em nonzero}~$w$, depending on the detailed structure of the upper
and lower Hubbard bands and the function $f(x)$, these two relations 
exclude each other. Therefore we obtain a contradiction 
already for some $U<U_c$, that is {\em in} the metallic regime.
This excludes solutions of the dynamical mean field equations describing 
the transition scenario~(\ref{scenario1}) within the skeleton 
expansion. 

In this context a comment on the frequently used  
``iterated perturbation theory'' (IPT) approach \cite{Georges92} 
to $d\rightarrow\infty$ problems seems in order. 
IPT is second order perturbation theory for the self--energy in~$U$
where the Weiss propagator~${\cal G}(\epsilon^+)$ is used for the
internal lines
\begin{eqnarray}
J^{\rm (IPT)}(\epsilon)&=&\pi U^2 \int_0^\epsilon d\mu
\,\rho_0(\epsilon-\mu) \nonumber \\
&&\times\int_0^\mu d\nu \,\rho_0(-\nu)\,\rho_0(\mu-\nu) 
\label{ipt}
\end{eqnarray}
with $\rho_0(\epsilon)=-\frac{1}{\pi}\mbox{Im}\, {\cal G}(\epsilon^+)$. 
This approximation leads to a metal--insulator transition that
is {\em not seen} when the full propagator is used for the internal
lines in similar schemes \cite{MuellerHartmann89}.
Since the non--interacting density of states $\rho_0(\epsilon)$ 
develops resonances on the energy 
scale~$\sqrt{w}t$ close to a
hypothetical metal--insulator transition like in Fig.~1, similar resonances
are found in $J^{\rm (IPT)}(\epsilon)$. 
This mechanism permits to fulfill the self--consistency
conditions within the IPT--approximation. However, from the results 
in this Letter it is clear that such resonances 
in $J^{\rm (IPT)}(\epsilon)$ are an artefact of the non--selfconsistent 
approximation used: If {\em all} orders of perturbation
theory are summed up there is no phase space for scattering
states available on the energy scale~$\sqrt{w}t$. This observation 
raises serious doubts whether IPT incorporates the correct mechanism 
that actually drives the transition. 

In summary in this Letter we have investigated the zero temperature 
metal--insulator transition in the half--filled Hubbard model
in the paramagnetic phase on a Bethe lattice with large 
connectivity \cite{lro}.
The framework used for this investigation was the skeleton expansion
to all orders: The pointwise convergence of
the skeleton expansion in the metallic phase is the basic 
{\em assumption} (but compare \cite{skeleton}) used in this Letter.
Therefore our analysis was restricted to the metallic
side of the transition since the skeleton expansion is known
not to converge for an insulator.

An important constraint for the transition scenario follows from
the asymptotic sum rule~(\ref{sumrule}) for the imaginary part of the 
self--energy. 
This sum rule leads to a {\em competition} between the availability
of phase space for scattering states and the suppression of 
spectral weight in the density of states in the vicinity of the
Fermi surface. 
This competition must be taken into account when investigating 
the Mott--Hubbard transition in the limit of large dimensions. 
It eliminates the scenario of a metal--insulator transition
with vanishing spectral weight in the vicinity of the Fermi surface,
compare Eq.~(\ref{scenario1}).
Only two transition scenarios are consistent with the skeleton expansion 
for $U<U_c$: (i)~A ``discontinuous'' transition occurs at $U=U_c$ in the 
sense that finite spectral weight is redistributed. 
(ii)~For~$U=U_c$ there is nonzero spectral weight in any neighborhood
of the Fermi surface (pseudogap--behavior), corresponding to some
non--Fermi liquid point separating metallic and insulating regimes.
Careful numerical studies are required to establish
whether such a pseudogap--solution of the self--consistency equations
is possible. Finally, the competition effect demonstrated above can be 
expected to be of importance for nonzero temperature too.

The author is greatly indebted to D.~Vollhardt, P.~G.~J.~van Dongen,
F.~Gebhard and W.~Metzner 
for valuable discussions and many important remarks.
Discussions on related subjects with W.~Hofstetter, J.~Schlipf and 
M.~Kollar are acknowledged. 



\begin{references}
\bibitem{lro} This is the usual setting in the literature
for investigating the Mott--Hubbard transition also
on other lattices. 
However, it should be kept in mind that the
$T=0$ Mott--Hubbard transition is generically hidden by 
an antiferromagnetic phase. For a critical discussion 
see e.g.\ Ref.~\protect\cite{Gebhard}. But notice that
the self--consistency condition (\protect\ref{Bethe_sc})
is identically also realized for a fully frustrated lattice
with random couplings~$t_{ij}$ where there is no such AF~phase
(for details see Ref.~\protect\cite{RMP96}). Instead of the 
Bethe lattice we can therefore also use this fully frustrated
model as the starting point.
\bibitem{Uc1Uc2} In the literature often two critical points 
$U_{c1}<U_{c2}$ are discussed \protect\cite{RMP96,Logan98}. 
This leads to a coexistence region $[U_{c1},U_{c2}]$ of an insulating
and a metallic solution. The idea is that at~$U_{c1}$ the insulating
gap and at~$U_{c2}$ the quasiparticle peak of the corresponding
branch of solution vanishes. Notice that
the discussion in this Letter always focusses on the metallic branch,
therefore $U_c$ as defined in this Letter corresponds to the
above~$U_{c2}$.
\bibitem{uniformity} This proportionality holds uniformly in~$n$,
i.e. the infrared limit is approached uniformly for all~$n$ 
as $w\rightarrow 0$. This is due to the separation of the
Hamiltonian in a gapped high--energy part and an effective 
low--energy Kondo Hamiltonian after a suitable unitary
transformation that becomes possible when (\protect\ref{scenario1})
holds \protect\cite{Moeller95}: The low--energy
Hamiltonian can be made dimensionless after dividing by~$wt$
and its dimensionless couplings remain essentially unchanged
as~$U\uparrow U_c$. Therefore all $\Gamma^{(n)}$ contain a trivial overall
$w^{-2n}$--dependence and remain otherwise form--invariant
functions of dimensionless variables $\xi_i/wt$ as
$U\uparrow U_c$. This unitary mapping to a well--understood
Kondo Hamiltonian also guarantees that all vertex functions 
in~(\protect\ref{skel_n})
are well--behaved in the Fermi liquid region of order~$wt$ 
around~$\epsilon_F$. Both these properties need no longer
hold for $U\uparrow U_c$ in a pseudogap--transition scenario with a
critical point at~$U_c$.
\bibitem{skeleton} Notice however that the existence of the skeleton
expansion is already needed for the derivation of the local nature
of perturbation theory in infinite dimensions along the lines
of Metzner and Vollhardt~\protect\cite{Metzner89}.
\end{references}
\end{document}